\documentstyle[12pt,epsf]{article}
\begin{document}

\title{Phase Structure\\
of SU(2) Lattice Gauge Theory\\
with Quantum Gravity\thanks{Submitted to Physics Letters B.}}

\author{Bernd A. Berg\thanks{On sabbatical from the
Florida State University.}\\
Wissenschaftskolleg zu Berlin\\
Wallotstrasse 19, W-1000 Berlin 33, Germany\\
and\\
Balasubramanian Krishnan\\
Department of Physics and\\
Supercomputer Computations Research Institute\\
Florida State University, Tallahassee, FL 32306, USA}

\maketitle

\begin{abstract}
We investigate 4$d$ SU(2) lattice gauge theory with Regge--Einstein
quantum gravity on a dynamically coupled Regge skeleton. To overview the
phase diagram we perform simulations on a small $2\cdot 4^3$
system. Evidence for an entropy--dominated disordered, an
entropy--dominated ordered and an ill--defined region is presented.
\end{abstract}
\newpage

In recent years various exploratory simulations of pure 4$d$ quantum
gravity have been carried out, see ref.~\cite{Me,Ha} for recent overviews. In
ref.~\cite{B1} one of the authors reported numerical evidence for an
``entropy--dominated'' phase, where the partition function is
well--defined, although the Regge-Einstein action is unbounded.
Essentially, these results were confirmed by Beirl et al.,\cite{Vienna},
where also some universality under changes of the measure was reported.
The question of possible physical relevance of the entropy-dominated
phase was first addressed in ref.~\cite{B2}, and it was suggested to show
that the entropy--dominated phase can accommodate hadronic masses
$m_{\rm h}$ many order of magnitude smaller than the Planck mass $m_{\rm P}$.
SU(2) lattice gauge theory was suggested to represent the matter
fields, because it stands out as the simplest 4$d$ asymptotically
free quantum field theory. In ref.~\cite{BKK}, see ref.~\cite{BK} for a concise
summary, a research program along this line of thought was outlined,
and supported by exploratory numerical results
from an $2\cdot 4^3$ lattice. For the coupled system the entropy
dominated phase survives, and within it an order--disorder transition
could be located. In a continuum or almost continuum limit, if it exists,
the transition will have the usual interpretation of a deconfining phase
transition, the ordered phase defines a quark--gluon plasma and in the
disordered phase quarks are confined. To provide modest evidence for
a consistent picture, one has to show that $m_{\rm h}/m_{\rm P}$ can
be decreased. This will require a finite size scaling analysis, and
simulations on $4\cdot 8^3$ and larger systems. They are out of reach
for present medium--sized computer systems. In contrast, a study of the
phase diagram in the ($m_{P}^{2}$, $\beta$) plane is feasible. To map
out the phase diagram we performed new simulations, which
are reported in this paper.

Let us recall euclidean quantum gravity on a Regge skeleton. The
Regge--Einstein action is given by
$$ S_{\rm RE} = 2 m^2_{\rm P} \sum_t \alpha_t A_t. \eqno(1)$$
Here $m_{\rm P}$ is the Planck mass, the sum is over all triangles
($d-2$ simplices) of a 4$d$ Regge skeleton, $A_t$ denotes the area of
the corresponding triangle and $\alpha_t$ its deficit angle. As in
previous work, our Regge skeleton will be given by the standard
tesselation of the hypercubical 4$d$ lattice. It is
well--known that the Regge--Einstein action is unbounded. Nevertheless
the theory could exist due to entropy effects. For volume
$V = {\rm const.}$, numerical evidence was reported in ref.~\cite{B1}
for an entropy--dominated region, such that the partition function
$$ Z = \int \prod_l {dx_l \over x_l}\ e^{S_{\rm RE}}  \eqno(2) $$
exists for $m^2_{\rm P} < m^2_{\rm MAX}$. Here $m^2_{\rm MAX}$ limits
the well--defined region. As in ref.~\cite{B1} we use the expectation value
$<v_p>$, the average volume of a 4-simplex, to define our
length unit as $l_0=<v_p>^{1/4}$. Then $m^2_{\rm MAX} \ge 0.02 l_0^{-2}$
is found. Henceforth, all quantities will be  expressed in units of
$l_0$, which will not be explicitly written down anymore.
In a fundamental length scenario\cite{TD1} $l_0$ stays finite
in physical units (for instance Fermi), whereas $l_0\to 0$ for a
continuum theory. A fundamental length theory is attractive under the
assumption that it will define a ``natural'' regularization scheme,
such that no renormalization occurs when physical quantities are
calculated in this scheme. If this is not the case, the continuum
scenario is preferred, as it offers the possibility that (after
renormalization) physical results will not depend on microscopic
details of the regularization, {\it i.e.} be universal in the usual
sense of lattice gauge theory. These two scenarios are presumably
indistinguishable in present day numerical simulations.
Classically our condition $V = {\rm const.}$ corresponds to a
cosmological constant
$$\lambda = - {1\over 2} m^2_{\rm P} V^{-1} \int d^4x \sqrt{g} R. \eqno(3)$$
This is seen by applying the variational principle to
$S = m^2_{\rm P} V^{-1/2} \int d^4x \sqrt{g} R$ and comparing the result
with the one obtained from
$S = m^2_{\rm P} \int d^4 x \sqrt{g} R + \lambda V$. Equation (3) is
consistent with the observed value $\lambda < 10^{-122} m^4_{\rm P}$,
and this would survive quantization if renormalization is not necessary.

Our complete action is
$$ S = S_{\rm RE} + S_{\rm gauge},  \eqno(4) $$
where
$$ S_{\rm gauge} = -{\beta \over 2} \sum_t W_t {\rm Re} [Tr(1-U_t)]. \eqno(5)$$
Here $U_t$ is the ordered product of SU(2) matrices attached to the links
of triangle $t$.
Coupling of the gauge and the gravity part is entirely achieved through
the dynamical weight factors $W_t$.
To obtain the correct classical limit, the weight factors
$W_t$ have to fulfill the constraint\cite{TD2}
$$ \sum_t W_t A_t^2 = {\rm const.} V, \eqno(6a)$$
which should be supplemented by ref.~\cite{BKK}
$$ W_t \ge 0 . \eqno(6b)$$
These conditions leave ample freedom for choice, and in the following
we use the barycentric implementation.\cite{BKK}
At $m^2_{\rm P}=0.005$ this system was investigated for various $\beta$
values. This is a data point well inside the
entropy--dominated region. In the present paper we report additional
results at $m^2_{\rm P} = 0.010$, 0.015, 0.020, 0.025 and 0.0375.

Let us first consider $m^2_{\rm P} = 0.010$, 0.015 and 0.020.
There is evidence for the disorder--order phase transition at all of
these Planck masses and Polyakov loop histograms for some relevant $\beta$
values are shown in figures 1, 2 and 3.  As in ref.~\cite{BKK} 20,000
thermalization sweeps were performed before measurement.
The subsequent statistics for each histogram is $\beta$ dependent
and lies in the range 20,000--60,000 sweeps. More complete numerical
results and additional data points are given in ref.~\cite{Ki}.
A three-peak structure is sometimes observed when
both confined and deconfined phases coexist, for example in figure~1.
This is suggestive of a first order transition in the vicinity
of the coupling. To substantiate this conjecture, would require
finite size scaling investigations (see for instance ref.~\cite{ABS})
on $2\cdot 6^3$, $2\cdot 8^3$, ... systems, and has to be postponed until
more powerful computational facilities are readily available.
As function of $m^2_{\rm P}$ the region of $\beta$ around which the
transition occurs does not vary much.  Tentatively, we identify it to
be between $\beta\,=1.46\,$ and $\beta\,=1.57\,$.

\begin{figure}
\makebox[3in][l]{
 \vbox to 3in{
  \vfill
\includegraphics{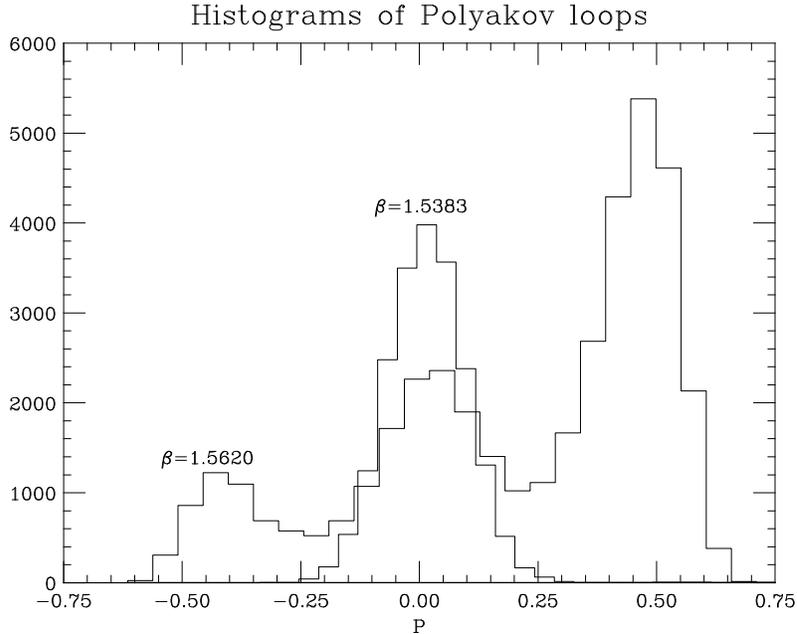}
    }
      \vspace{-\baselineskip}
      }
\vspace{10pt}
\caption{Histograms of Polyakov loops at two different gauge couplings
         for Planck mass $m_{P}^{2}\,=\,0.010$.}
\end{figure}
\begin{figure}
\makebox[3in][l]{
 \vbox to 3in{
  \vfill
\includegraphics{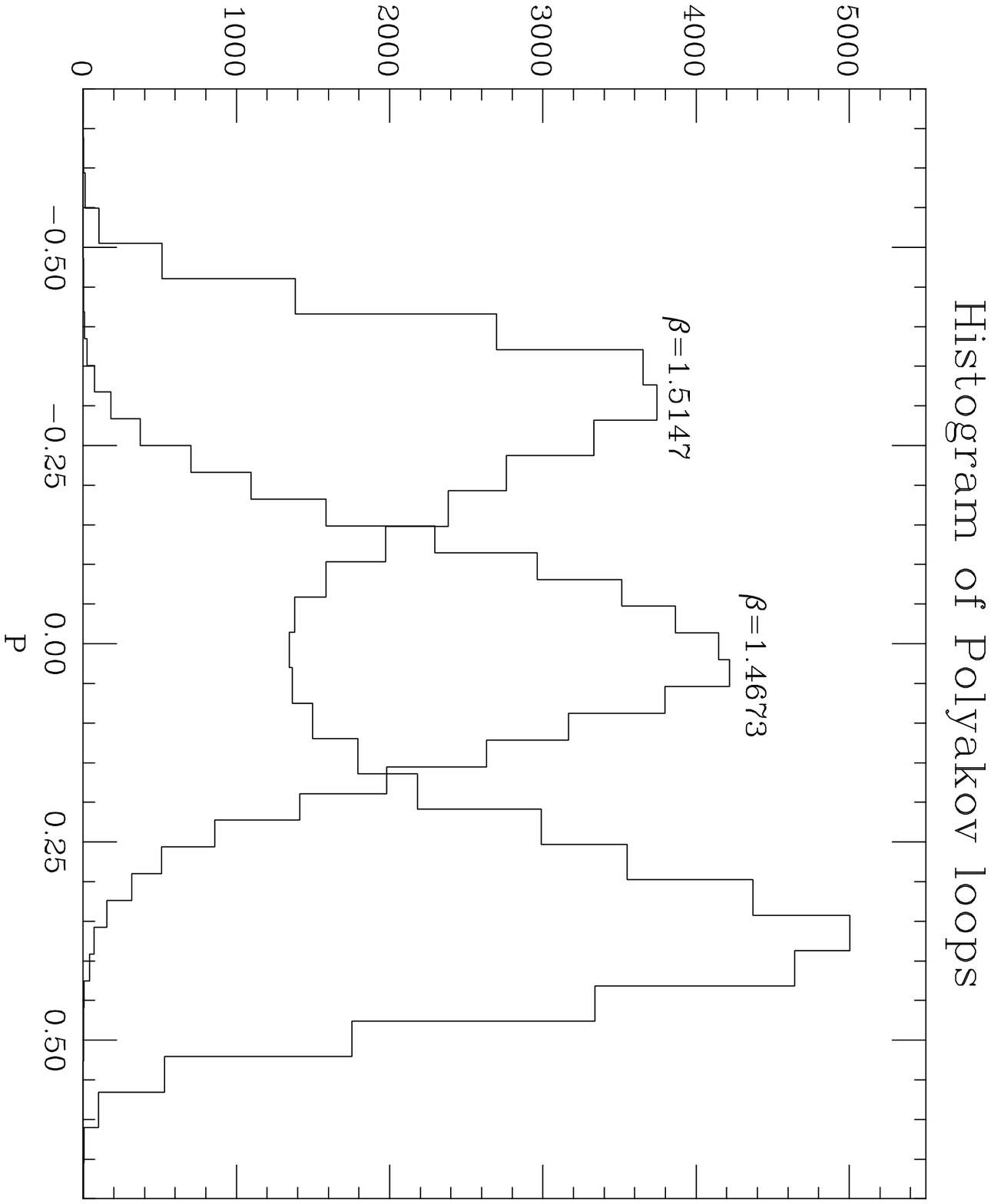}
    }
      \vspace{-\baselineskip}
      }
\vspace{10pt}
\caption{Histograms of Polyakov loops at two different gauge couplings
         for Planck mass $m_{P}^{2}\,=\,0.015$.}
\end{figure}
\begin{figure}
\makebox[3in][l]{
 \vbox to 3in{
  \vfill
\includegraphics{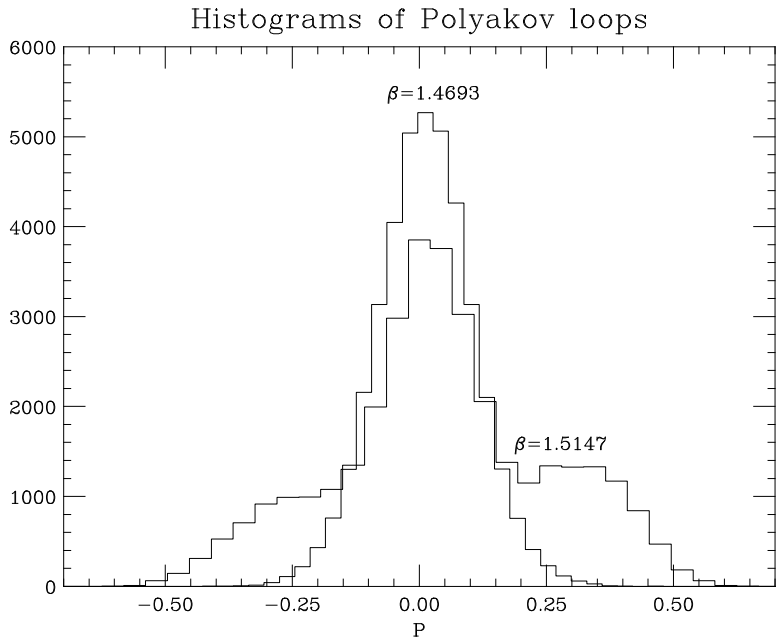}
    }
      \vspace{-\baselineskip}
      }
\vspace{10pt}
\caption{Histograms of Polyakov loops at two different gauge couplings
for Planck mass $m_{P}^{2}\,=\,0.020$.}
\end{figure}

The next step is to proceed towards higher Planck masses. At a high
enough value of the Planck mass one does expect that the system will find
itself in unstable states, similar as in the pure gravity
case\cite{B1,Vienna}. Indeed, suggestions to that effect begin to appear at
$m_{P}^{2}\,=\,0.025$ and $\beta\,=\,1.42$.  Figure~4 shows the time
history of the gravity action $S_{\rm RE}$. A  two-peak structure, seen usually
in first-order phase transitions, develops. However, it is unclear whether
one can characterize the behavior as such because, as in the
pure gravity case, the partition function likely ceases to exist when
one is outside the entropy-dominated region. In any case the two peak
structure characterizes competing effects of the entropy versus the
unboundedness of the Regge-Einstein action.

\begin{figure}
\makebox[3in][l]{
 \vbox to 3in{
  \vfill
\includegraphics{simfig12.ps}
    }
      \vspace{-\baselineskip}
      }
\vspace{10pt}
\caption{Time history of the  gravity action $S_{RE}$ for
$m_{P}^{2}\,=\,0.025$ and $\beta\,=\,1.42$. Each point is an average
over ten consecutive measurements.}
\end{figure}

At Planck mass $0.025$ we seem to have reached the region where entropy
effects from the measure begin to lose control over the unboundedness of
the action. In figure~6 we present the time history of the gravity action
at an even higher Planck mass $0.0375$ and $\beta\,=\,1.42$. It is
constantly on the rise and the system is clearly unstable.
\begin{figure}
\makebox[3in][l]{
 \vbox to 3in{
  \vfill
\includegraphics{simfig17.ps}
    }
\vspace{-\baselineskip} }
\vspace{10pt}
\caption{Time history of the gravity action $S_{RE}$ for
$m_{P}^{2}\,=\,0.0375$ and $\beta\,=\,1.42$.
Each point is an average over ten consecutive measurements.}
\end{figure}

The ($m_{P}^{2}\,,\,\beta$) phase diagram is summarized
in figure~6. Regions A and C constitute the entropy-dominated region
which is estimated to be $0\,\le\,m^2_{\rm P}\,\le\,0.02$,
possibly up to $m^2_{\rm P}\,\le\,0.03$. In region B the partition
function ceases to exist and is assumed to be meaningless.
Region C is the ordered phase where, in the infinite volume limit,
Polyakov loops assume non-zero expectation values and region A is the
disordered phase where the expectation value of the Polyakov loops are zero.
In the present paper the hadronic mass (in our unit $l_0^{-1}$) is given
by the ``deconfining'' temperature
$$ T_c = {1\over 2<x_l>_0}, \eqno(7) $$
where $<x_l>_0$ is the expectation value of the link length in the
short ($L_0=2$) direction of the hypercubic lattice. Fairly independent
of $m_{\rm P}$ this value stays around $<x_l>_0\,\approx 3$.\cite{Ki}
Assuming $m^2_{\rm MAX}=0.025$, one obtains then
$m_{\rm P}/m_{\rm h} \le 0.95$ for the $2\cdot 4^3$ lattice. With
increased computational power one may first envision a finite size scaling
analysis of $2\cdot L^3$ systems towards $L\to\infty$, to confirm and
significantly improve the accuracy of the phase diagram of figure~6.
Simultaneously, one should study an $4\cdot 8^3$ lattice on the present
heuristic level, to obtain evidence that the ratio $m_{\rm P}/m_{\rm h}$
can indeed be increased (naively by a factor of two).

\begin{figure}
\makebox[3in][l]{
  \vbox to 3in{
     \vfill
      \includegraphics{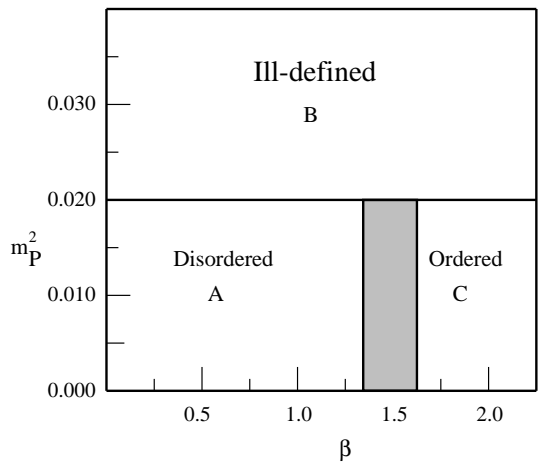}
    }
      \vspace{-\baselineskip} }
\caption{The phase diagram of the system, obtained from simulations on
a $2\cdot4^{3}$ system.}
\end{figure}

\section*{Acknowledgements}
One of the authors(BB) likes to thank Wolfgang
Beirl for pointing out equation (3). This research project was partially
funded by the Department of Energy under contracts DE-FG05-87ER40139
and DE-FC05-85ER2500.

\newpage
\bibliographystyle{unsrt}

\end{document}